\documentclass[a4paper,nodate]{sab} 
\usepackage{graphicx} 
\usepackage{txfonts,natbib} 
\bibpunct{(}{)}{;}{a}{}{,} 
\usepackage{hyperref}
\usepackage{url}
\usepackage[T1]{fontenc} 
\usepackage{t1enc} 
\usepackage{ulem}
\usepackage{xcolor}
\renewcommand\ackname{Acknowledgements.} 
\renewcommand\andname{\&}
\renewcommand\keywordname{Keywords.} 
\renewcommand\figureptname{Figure}
\renewcommand\tableptname{Table}

\renewcommand\noteaddname{Note added to proofs.}
\renewcommand\andname{\&}

\idline{Boletim da Sociedade Astron\^omica Brasileira, {\bf 31}, no. 1, XX-XX}{3}

\def\feii{{Fe {\sc ii}}}

\def\kms{km~s$^{-1}$}

\def\rfe{R$_\mathrm{FeII}$}
\def\hb{H$\beta$}

\def\rblr{R$_{\rm BLR}$}
\def\lopt{L$_{\rm 5100}$}

\def\zsun{Z$_\odot$\/}

\begin{document} 
\title{Saturation of the curve} 
\subtitle{diagnostics of the continuum and \hb{} emission in Population B active galaxy NGC 5548} 

\author{S. Panda \inst{1}, E. Bon \inst{2}, P. Marziani \inst{3} \and N. Bon \inst{2}} 
\institute{Laborat\'orio Nacional de Astrof\'isica - Rua dos Estados Unidos 154, Bairro das Na\c c\~oes. CEP 37504-364, Itajub\'a, MG, Brazil. \email{spanda@lna.br} \and Astronomical Observatory Belgrade, Volgina 7, 11060 Belgrade, Serbia. \and INAF-Astronomical Observatory of Padova, Vicolo dell'Osservatorio, 5, 35122 Padova PD, Italy.}
\date{Received } 

\Abstract {NGC 5548 has been hailed as an archetypical type-1 active galactic nuclei (AGN) and serves as a valuable laboratory to study the long-term variation of its broad-line region. In this work, we re-affirm our finding on the connection between the continuum variability in the optical regime and the corresponding \hb{} response to it, in order to realize the increase, albeit with a gradual saturation, in the \hb{} emitting luminosity with increasing AGN continuum. This effect is also known as the Pronik-Chuvaev effect after the authors who first demonstrated this effect using long-term monitoring of another well-studied Type-1 AGN - Mrk 6. We employ a homogeneous, multi-component spectral fitting procedure over a broad range of spectral epochs that is then used to create the continuum and \hb{} light curves. We focus on the epoch range 48636-49686 MJD, different from our previous analysis. We again notice a clear signature of shallowing in the trend between the \hb{} and the continuum luminosities. We attempt to recover this \hb{} emission trend as a response to a significant continuum flux increase using CLOUDY photoionization simulations and employ a suitable broad-band spectral energy distribution (SED) for this source. We explore the wide range in the physical parameters space for modelling the \hb{} emission from the broad-line region (BLR) appropriate for this source. We employ a constant density, single cloud model approach in this study and successfully recover the observed shallowing of the \hb{} emission with respect to the rising AGN continuum. In addition, we are able to provide constraints on the local BLR densities and the location of the \hb{} emitting BLR, the latter agreeing with the \hb{} time-lags reported from the long-term reverberation mapping monitoring for this source. On the contrary, we do not find a significant breathing effect in the location of the \hb{} line-emitting BLR for this epoch in NGC 5548.}{NGC 5548 foi definido como um arquetípico núcleo galáctico ativo tipo 1 (AGN) e serve como um laboratório valioso para estudar a variação de longo prazo de sua região de linha ampla. Neste trabalho, reafirmamos nossa constatação sobre a conexão entre a variabilidade do contínuo no regime óptico e a correspondente resposta \hb{} a ele, a fim de perceber o aumento, ainda que com saturação gradual, no \hb{ } emitindo luminosidade com o contínuo AGN crescente. Esse efeito também é conhecido como efeito Pronik-Chuvaev após os autores que demonstraram esse efeito pela primeira vez usando o monitoramento de longo prazo de outro AGN Tipo 1 bem estudado - Mrk 6. Empregamos um procedimento de ajuste espectral homogêneo e multicomponente sobre um ampla gama de épocas espectrais que é então usada para criar as curvas de luz contínuas e \hb{}. Focamos na faixa de época 48636-49686 MJD, diferente de nossa análise anterior. Novamente notamos uma assinatura clara de redução na tendência entre o \hb{} e as luminosidades contínuas. Tentamos recuperar essa tendência de emissão \hb{} como resposta a um aumento significativo de fluxo contínuo usando simulações de fotoionização CLOUDY e empregamos uma distribuição de energia espectral de banda larga (SED) adequada para essa fonte. Exploramos a ampla faixa no espaço de parâmetros físicos para modelar a emissão \hb{} da região de linha ampla (BLR) apropriada para essa fonte. Empregamos uma abordagem de modelo de nuvem única de densidade constante neste estudo e recuperamos com sucesso a redução observada da emissão \hb{} em relação ao contínuo crescente de AGN. Além disso, podemos fornecer restrições sobre as densidades de BLR locais e a localização do \hb{} que emite BLR, este último concordando com os \hb{} defasagens de tempo relatadas do monitoramento de mapeamento de reverberação de longo prazo para essa fonte . Pelo contrário, não encontramos um efeito respiratório significativo na localização do BLR emissor de linha \hb{} para esta época em NGC 5548.}


\maketitle 

\section{Introduction}

NGC 5548 is well-known as an archetypical Type-1 AGN \citep{ursinietal15} and is a valuable laboratory to study the long-term variation of the broad-line region \citep[BLR][]{wanders1996,sergeev2007}. It has been the target of many reverberation campaigns, notable among them were the International AGN Watch and AGN STORM programs \citep[see][]{Petersonetal2002,derosaetal2015}. These campaigns have successively provided information on the geometry, ionization structure, and kinematics of the broad-line emitting region in this source \citep[][and references therein]{derosaetal2015,horneetal2021}.

\begin{figure}
    \centering
    \includegraphics[width=0.45\textwidth]{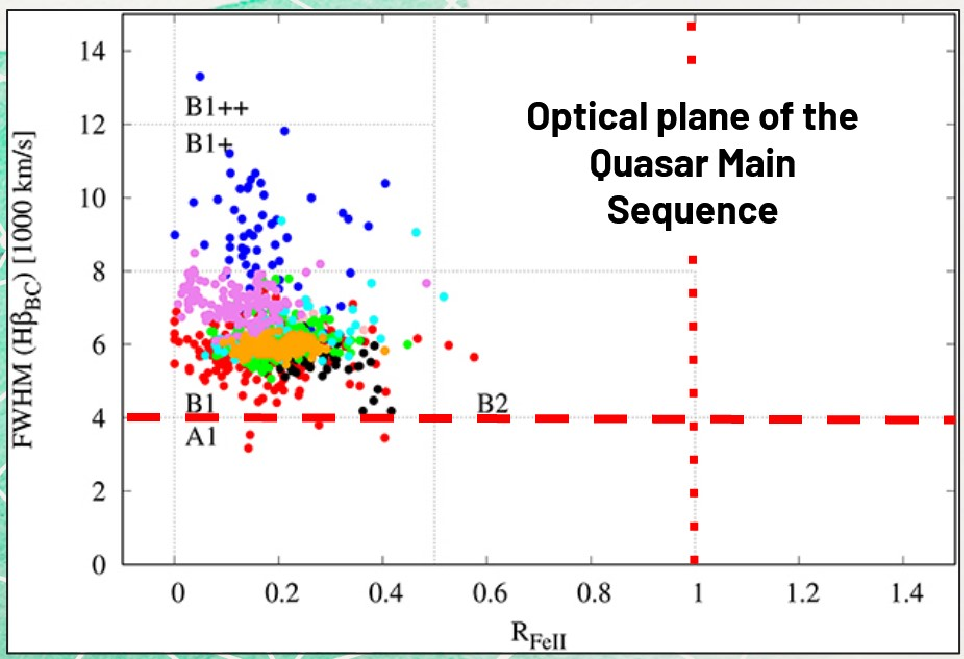}
    \caption{The Eigenvector 1 diagram of NGC 5548 spectral properties during 43 years of monitoring campaigns. Colours correspond to different time intervals. The vertical dotted line represents the \rfe{} = 1 that classifies sources either into low or high \feii{}-emitters. The horizontal dashed line at FWHM(\hb{}) = 4000 \kms{} demarcates the Population A sources from those that belong to Population B.  Abridged version of the original from \citet{bonetal2018}.}
    \label{fig:qms}
\end{figure}

\begin{figure}
    \centering
    \includegraphics[width=0.45\textwidth]{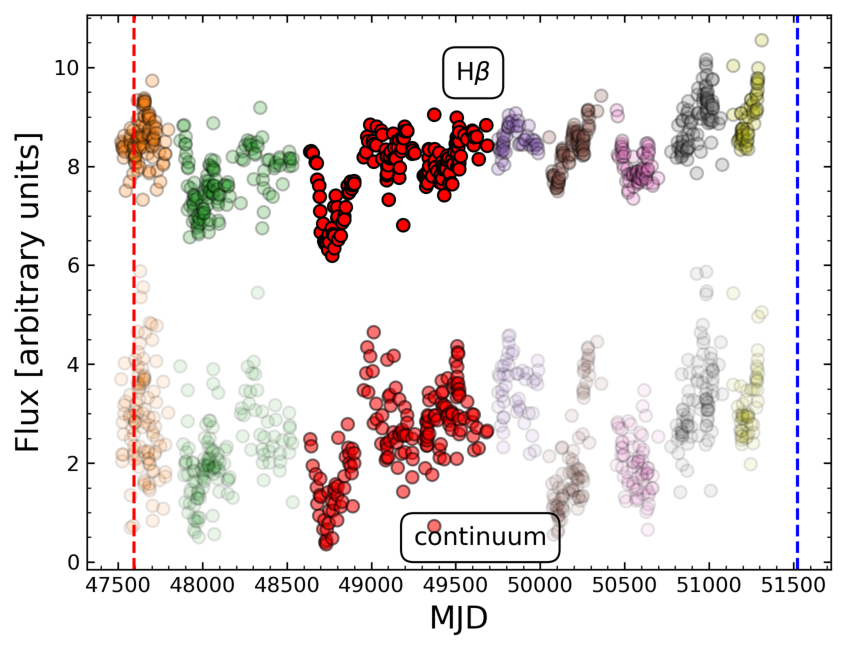}
    \caption{\hb{} flux and 5100\AA\ continuum (\lopt{}) flux light curves for NGC 5548 extracted from spectroscopic multi-component fitting. The range between 48636-49686 MJD used in this work is highlighted in red for both the \hb{} and \lopt{}. The red \citep{Clavel_1991ApJ...366...64C} and blue \citep{Kaastra_2002A&A...386..427K} dashed lines represent the epochs for the multi-wavelength SEDs available for this source in the literature in this time range.}
    \label{fig:lc}
\end{figure}

A different yet useful way to look at the evolution of this source is tracing it along the optical plane of the Eigenvector 1 schema \citep[][see also Figure \ref{fig:qms}]{bg92,bonetal2018, Panda_CLAGN_2022arXiv220610056P}. The Eigenvector 1  (a.k.a as the optical plane of the main sequence of quasars)  is often represented as a correlation in the 2D observational space defined by the FWHM of broad \hb{} emission line and the strength of the optical \feii{} emission\footnote{the \feii{} emission is the integrated emission within 4434-4684 \AA.} with respect to \hb{} \citep[][and references therein]{bg92,sulenticetal00c,sh14,sul15,mar18,panda18b,panda19b}.  Within this optical plane, NGC 5548 appears as a prototypical Population B source \citep{sulenticetal00c}, where Population B includes sources radiating at modest Eddington ratio ($\lesssim 0.2$) with FWHM(\hb{}) $\gtrsim$ 4000 \kms{}. NGC 5548 shows significant variability in the optical and UV bands of the electromagnetic spectrum and has been the target of over a dozen reverberation mapping campaigns \citep[see][and references therein]{Luetal2016}. In addition to a prototypical spectrum, the source even with transitions from low to high states stays within the Pop. B class in terms of FWHM \hb\ and \feii{}  \citep{bonetal2018}.

In this paper, we further the recent results that were published in \citet{Panda_2022AN....34310091P} wherein we test the connection between the continuum variability in the optical regime and the corresponding \hb{} response to it for NGC 5548, in order to realize the increase, albeit with a gradual saturation, in the \hb{} emitting luminosity with increasing continuum. This effect, also known as the Pronik-Chuvaev effect after the authors who first demonstrated this effect using long-term monitoring of Mrk 6 \citep{pronik_chuvaev1972}, has been also observed and studied in a handful of nearby sources, e.g. NGC 4051 and NGC 4151 \citep[][and references therein]{Wang2005, Shapavalova2008,gaskell2021}. With our work, we are now able to quantify this saturation and tie it to the changes in the physical parameters leading to the emission of the \hb{} in the BLR. We briefly outline the multi-component spectral fitting procedure that is then used to create the continuum and \hb{} light curves over a broad time range\footnote{from these long-term light curves, we focus on the trend for the epoch range between 48636-49686 MJD. This is different from the epoch range analyzed in our previous work, i.e., 51170-52174 MJD.} and describes the photoionization modelling setup involving a carefully, observation-oriented parameter space study that includes accounting for the continuum source and the BLR physical conditions using constant density single cloud models. Next, we describe our results confirming yet again the saturation of \hb{} with increasing continuum luminosity that can be successfully reproduced with our photoionization modelling. This allows us to provide constraints on the BLR local densities and the location of the \hb{} emitting BLR, the latter agreeing with the \hb{} time-lags reported from the long-term reverberation mapping monitoring for this source. Throughout this work, we assume a standard cosmological model with $\Omega_{\Lambda}$ = 0.7, $\Omega_{m}$ = 0.3, and H$_0$ = 70 \kms{} Mpc$^{-1}$.

\section{The gradual flattening in the \hb{} with increasing AGN continuum}

We recover the information of the continuum (at 5100\AA) and the \hb{} emission line from the long-term monitoring of NGC 5548. 
%\sout{Contrary to} 
At variance with our previous work \citep{Panda_2022AN....34310091P}, we focus on the observational light curves between the time range (in Modified Julian Days) 48636-49686 (see the highlighted regions in red in Figure \ref{fig:lc}). The choice of the time range is supported by the broad range covered by the \hb{} and the continuum luminosity accompanied by the high level of variability in the two light curves. We use the publicly available spectroscopic observations to measure the \hb{} and continuum at 5100\AA~ fluxes.

\begin{figure}
    \centering
    \includegraphics[width=0.45\textwidth]{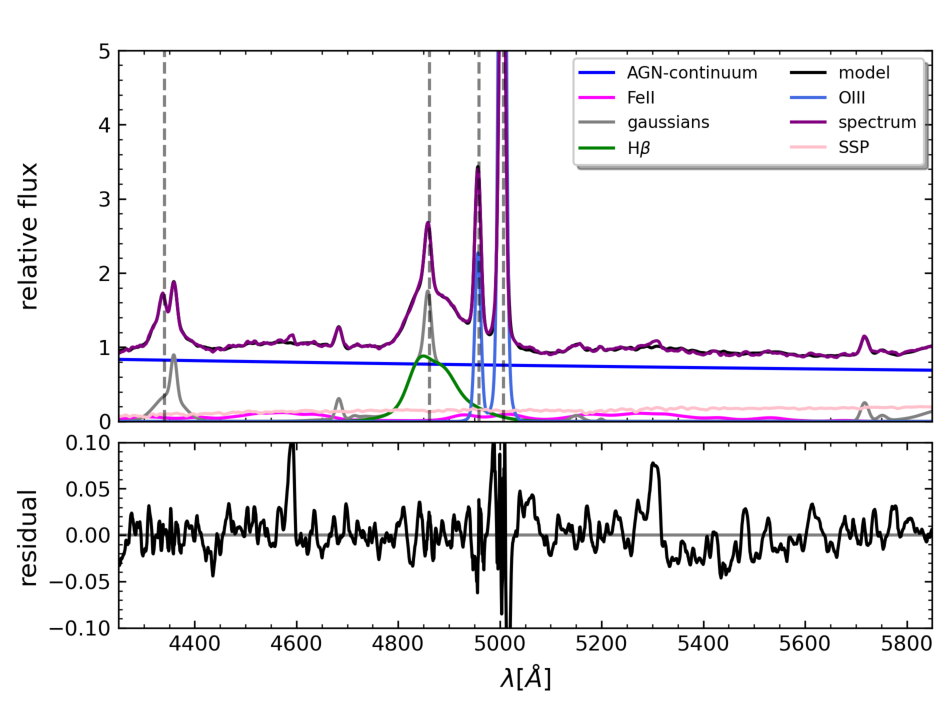}
    \caption{Spectral decomposition shown for a representative spectrum for NGC 5548 used in this work. The lower panel shows the residual for the observed spectrum (in purple) and the modelled spectrum (in black) over the wavelength range. The various components depicted in the figure are AGN power-law continuum (in dark blue), \feii{} pseudo-continuum (in magenta), extra Gaussians (in grey), \hb{} (in green), overall model (in black), [O III] doublet (in blue), observed spectrum (in violet), and modelled host galaxy contribution (in pink).}
    \label{fig:spectrum}
\end{figure}

The spectra were compiled using AGN Watch \cite[see,][and the references within]{Petersonetal2002}. Spectral decomposition is performed using ULySS software\footnote{available at \href{http://ulyss.univ-lyon1.fr}{http://ulyss.univ-lyon1.fr}}, full spectrum fitting technique \citep{Koleva2009,Bon14,Bonetal20}. For Type-1 Seyfert spectral analysis, the code includes - power law AGN continuum, nebular continuum, emission lines and \feii{} pseudo-continuum that are fitted simultaneously in order to minimize the effects of dependencies between parameters of the model \citep[for more details see,][]{Bon14,Bonetal20}. The host galaxy parameters were obtained using the low-state spectra, where the contribution of the host absorption lines is more prominent, and these parameters were assumed to be the same for the rest of the monitoring spectra. The broad emission lines were modelled assuming several Gaussian components: two broad and one very broad. A representative spectral decomposition of and around the \hb{} line complex is shown in Figure \ref{fig:spectrum}. The details of the spectral decomposition are presented in \cite{Bon14,bonetal2016,bonetal2018,Bonetal20, Panda_2022AN....34310091P}.

\begin{figure}
\centering
\includegraphics[width=0.45\textwidth]{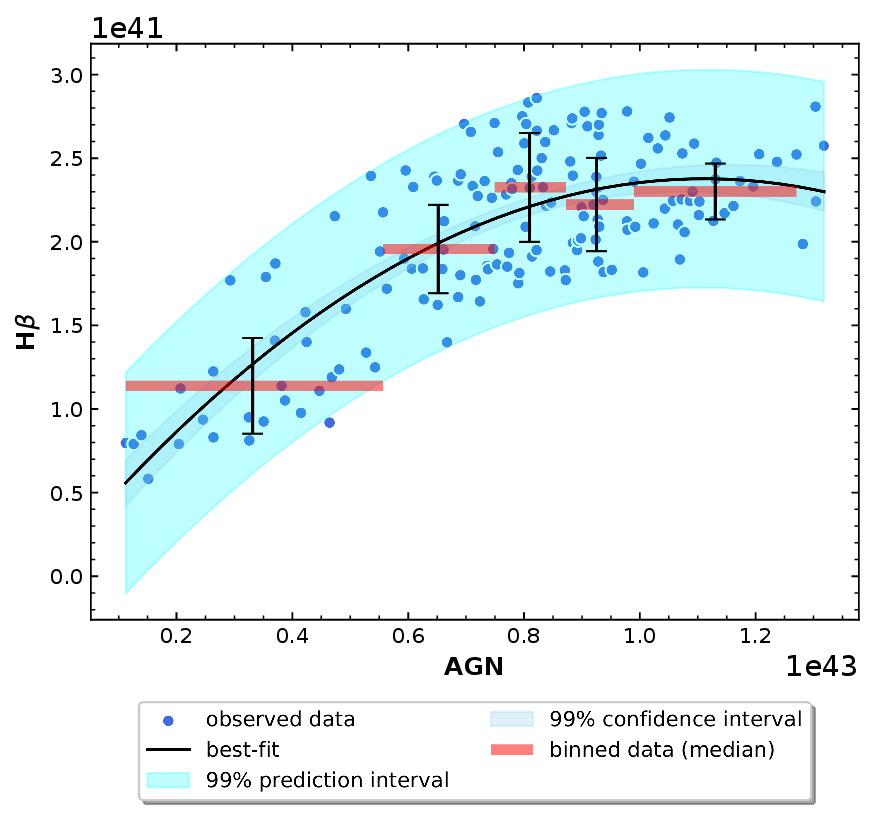}
\caption{Luminosity-luminosity plot between continuum
at 5100\AA\ vs \hb{} emission for the time range (in MJD) 48636 - 49686. The 159 data points (blue dots) are fitted with a 2$^{\rm nd}$ order polynomial regression fit (black solid curve). 99\% confidence and prediction bands are shown (cyan and light-blue shaded regions, respectively) generated from bootstrapping 1000 realizations of the data. The data is binned equally in 5 intervals (the extent of the bins is shown with horizontal red bars) to highlight the shallowing of the trend. The vertical error bars (in black) mark the interquartile range for the corresponding bins.}
    \label{fig:corr}
\end{figure}

The observed luminosity-luminosity correlation between \hb\ and the continuum is shown in Figure \ref{fig:corr}. To make this figure, we convert the fluxes into corresponding luminosities (in erg s$^{-1}$) for the epoch range 48636-49686 MJD which gives us 163 data points, identify and remove `real' outliers (4 data points) using their modified z-score\footnote{Modified z-score = 0.6745($x_{i}$ - $\tilde{x}$)/MAD, where $x_{i}$ represent each data value in the dataset, $\tilde{x}$ is the sample median, and MAD is the median absolute deviation of the dataset.}, and fit the observed trend with a 2$^{\rm{nd}}$ order polynomial regression fit of the form:

\begin{equation}
    {\rm{L_{H\beta}}} \;[\mathrm{erg\, s}^{-1}] = 1.2604 \cdot 10^{40} \,+\,0.0405\,{\rm L_{5100}} \,-\, 1.8210 \cdot 10^{-45}\,{\rm L_{5100}^2}.
\end{equation}

The Spearman's rank-correlation coefficient is\ $\rho$ = 0.5720, and the corresponding $p$-value is $\approx 3.55 \times 10^{-15}$. The best fit is supplemented with 99\%\ confidence and prediction intervals generated from bootstrapping 1,000 realizations of the observed data values. The trend obtained from this epoch range is relatively stronger ($\sim$20\% in the first-order slope) than our previous result for a different epoch range \citep{Panda_2022AN....34310091P}. This can be attributed to the different behaviour of the light curves in the two epoch ranges - in our earlier work, the used epoch range began at the peak of a flare and a declining feature thereafter (see Figure 2 in \citet{Panda_2022AN....34310091P}), while in this case, we have a rather complex behaviour. The two light curves begin at an average flux value and pass through a deep minimum (this is the lowest flux obtained in the full range shown in Figure \ref{fig:lc}) followed by a recovery to the initial `average' level and subsequent minor fluctuations about this mean value. 

To appreciate the flattening effect better, we bin the dataset in 5 bins that contain an equal number of data values. These bins are overplotted on the 2$^{\rm{nd}}$ order polynomial regression fit and are shown using red horizontal bars in Figure \ref{fig:corr}. To assess the dispersion in each bin, we show the interquartile range (IQR) using the black error bars. Immediately we can see two things - (i) the binned results are in good agreement with the 2$^{\rm{nd}}$ order polynomial regression fit, and (ii) the dispersion at the high state (at the maximum values for the continuum luminosities) is the smallest. We will later make use of these median bins and their IQRs to perform and compare with our photoionization models.

\section{Inferences from photoionization modelling}

\begin{figure}
    \centering
    \includegraphics[width=0.45\textwidth]{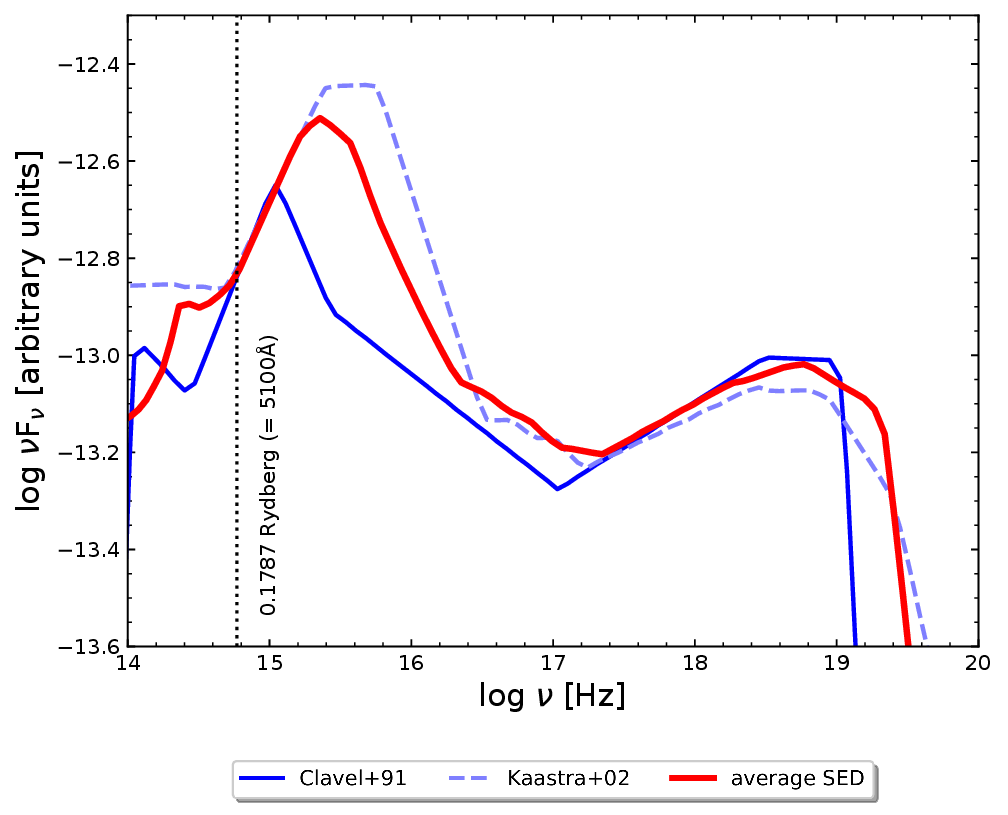}
    \caption{Observed multi-wavelength SEDs for NGC 5548: (a) for an earlier epoch (47593 MJD, \citet{Clavel_1991ApJ...366...64C} shown in solid blue), and (b) for a later epoch (51523 MJD, \citet{Kaastra_2002A&A...386..427K} shown in dashed blue). The average SED constructed using these two observed SEDs is shown in red.}
    \label{fig:sed}
\end{figure}

In order to perform the photoionization models, we first prepare an appropriate SED that is suitable for the epoch range that was analyzed in this work. From Figure \ref{fig:lc}, we know that the two available multi-wavelength SEDs from the literature are either suitable for an earlier (red dashed line, \citet{Clavel_1991ApJ...366...64C}) or for a later epoch range (blue dashed line, \citet{Kaastra_2002A&A...386..427K}). The two SEDs are shown in Figure \ref{fig:sed} using solid blue and dashed blue distributions, respectively, for the earlier and the later epoch ranges. Clearly, the \citet{Clavel_1991ApJ...366...64C} SED has a less prominent Big Blue Bump \citep{ss73,czerny87}, whereas the \citet{Kaastra_2002A&A...386..427K} SED has a considerable flux in this regime. To visualize this better, we normalize the SEDs at 5100\AA~. This allows us to assess the features of the two SEDs, especially around the 1 Rydberg energy region\footnote{1 Rydberg is equal to a frequency value 15.517 Hz (in log-scale).} which is roughly at the peak of the Kaastra et al.'s SED. The high-energy bump in the X-rays is roughly similar although Kaastra et al.'s SED has a weaker contribution relative to Clavel et al's. From the lightcurves, we notice that Clavel et al.'s epoch range corresponds to a low optical flux state, while Kaastra et al.'s epoch range is closer to a maximum flux value (see Figure \ref{fig:lc}). On the contrary, the epoch range we have analyzed in this work lies in between these two extremes. Thus, we perform an average of the two observed SEDs which is shown using the red distribution in Figure \ref{fig:sed} and incorporate this synthetic SED in our photoionization analyses.

We perform a suite of {\sc CLOUDY} \citep{f17} photoionization simulations considering a grid of single cloud models covering a wide range in the parameter space that includes (a) the continuum luminosity, (b) the distance between the continuum source and the onset of the BLR cloud (\rblr{}), (c) shape of the ionizing continuum characterized by the spectral energy distribution (SED) appropriate for the studied epoch for NGC 5548, (d) the BLR local density, and (e) the geometrical depth of the BLR clouds parameterized by the cloud column density. The first two parameters are constrained from prior observations, e.g., we consider the five luminosity states that are obtained by median-binning the observed data (see the horizontal red bars in Figure \ref{fig:corr}). These luminosity states are then provided as one of the inputs to the set of simulations. The radial distance between the source and the BLR cloud (\rblr{}) is set by the distance estimated from the time lags between the continuum and emission line light curves obtained from prior reverberation mapping studies carried out for NGC 5548 \citep[][and references therein]{horneetal2021} which follows the prescription used in our previous work \citep{Panda_2022AN....34310091P}. \citet{horneetal2021} estimated the range for this \hb{}-based \rblr{} to be as short as 5 light days\footnote{1 light day = 2.59$\times 10^{15}$ cm.} and extending up to 40 light days. The local BLR density is considered within the range $10^{9} \leq n_{\rm{H}} \leq 10^{13}\;(\rm{cm^{-3}}$) suitable for recovering the low-ionization emission (here, \hb{}) pertaining to the BLR \citep{panda_etal2020, korista_goad_2000}. We assume two values for the cloud column density, N$_{\rm H}$ = $10^{22}$ cm$^{-2}$ and $10^{23}$ cm$^{-2}$ and consider solar abundances (Z = \zsun{}) for the composition of the BLR cloud. This assumption of the composition is corroborated by previous studies for NGC 5548 \citep{dehghanianetal2019, Panda_2022AN....34310091P} and is thought to be usually appropriate for sources like NGC 5548 that belong to Population B \citep{punslyetal18a, panda19b}. The modelled luminosities for the \hb{} are extracted from these photoionization models.

\begin{figure*}[!htb]
    \centering
    \includegraphics[width=\textwidth]{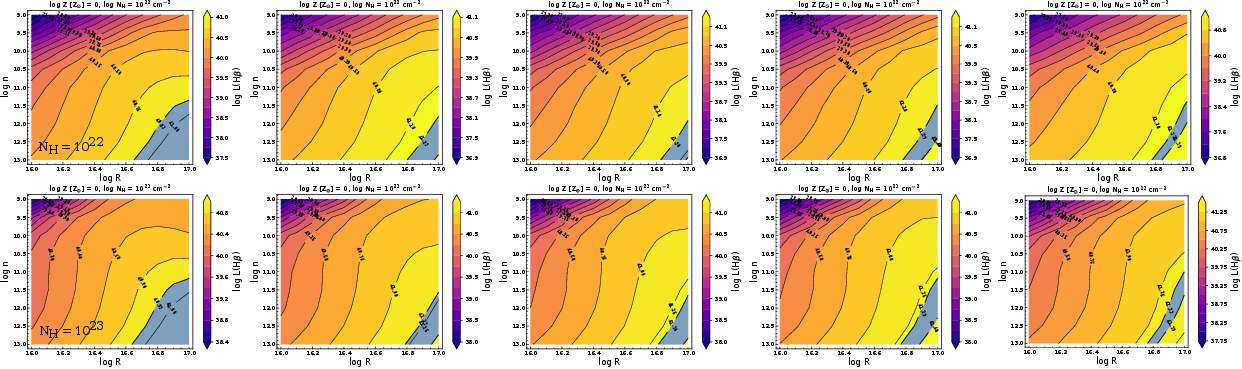}
    \caption{log \rblr{} - log n$_{\rm H}$ histogram matrix generated using the {\sc CLOUDY} simulations. The auxiliary axis is shown as a colourmap representing the luminosity of the \hb{} emission line (in log-scale) which is also represented using black contours in each panel. A covering factor of 20\% is used to re-scale the luminosities. The panels (from LEFT to RIGHT) represent increasing continuum/incident luminosity (in erg s$^{-1}$, in log-scale): 42.52, 42.81, 42.91, 42.97, and 43.05. A column density (N$_{\rm H}$) of 10$^{22}$ cm$^{-2}$ (top panels) and 10$^{23}$ cm$^{-2}$ (bottom panels), and solar abundances (Z = \zsun{}) are assumed for these photoionization models. The blue-shaded region highlights the observed IQRs corresponding to the 5 luminosity bins shown in Figure \ref{fig:corr} in each panel.}
    \label{fig:models}
\end{figure*}

The full, modelled parameter space considered in our photoionization computations is shown in Figure \ref{fig:models}. {\sc CLOUDY}, by default, provides the luminosities assuming a 100\% covering factor, i.e. solid angle that is equal to 4$\pi$ steradians. We utilize a covering factor of 20\% that is found to be viable for a near-Keplerian distribution of the BLR clouds \citep{korista_goad_2000, baldwin_etal2004, sarkar_etal2021, panda2021}. In order to assess the performance of the photoionization computations against the observed values obtained from the \textit{flattening} curve, we supplement the 2D histograms with the corresponding estimation of the \hb{} line luminosities that are shown using the blue shaded regions in each panel of Figure \ref{fig:models}. The extent of the shaded region is set by the corresponding inter-quartile range as shown on the \textit{flattening} curve, i.e. Figure \ref{fig:corr}. In both cases of the column densities, we see a gradual shrinking\footnote{The tapering of the width of the shaded region as we increase the continuum luminosity is due to the lowering of the dispersion in the corresponding luminosity bin that is highlighted by the smaller inter-quartile range.} of the overlapping region between the observed range and the modelled grid, as we go from the lowest AGN continuum luminosity state (log L$_{\rm 5100}$ = 42.52, in erg s$^{-1}$) to the highest state (log L$_{\rm 5100}$ = 43.05, in erg s$^{-1}$). Hence, contrary to our previous results, both the column densities give comparable constraints on the location and local density of the \hb{} emitting region.

\begin{figure}[!htb]
    \centering
    \includegraphics[width=0.85\columnwidth]{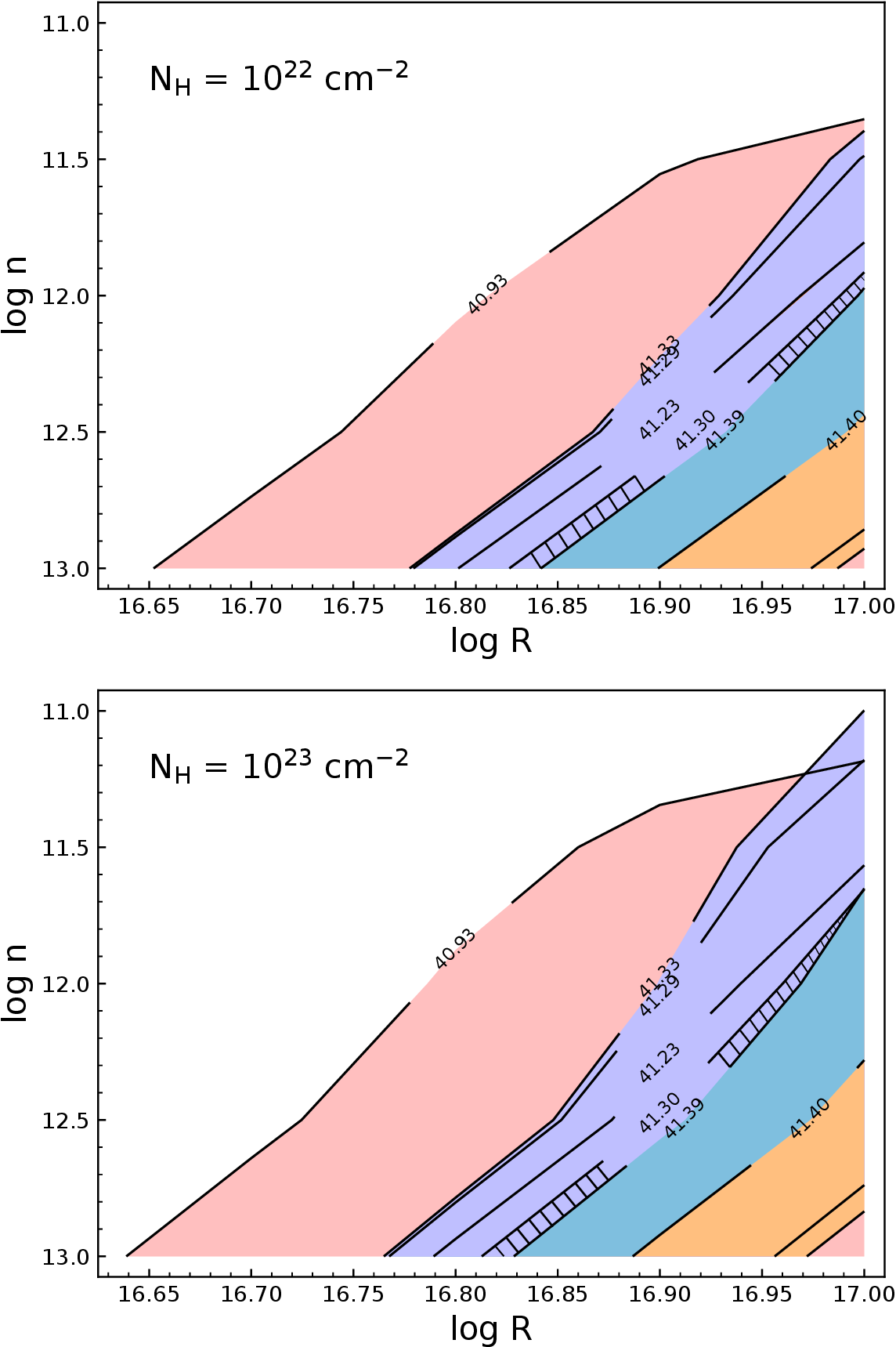}
    \caption{Overlapping regions in each of the panels shown in Figure \ref{fig:models} showing the agreement between the observed range and modelled grid in the log \rblr{} - log n$_{\rm H}$ parameter space. The upper panel shows the extent of overlap for the column density, 10$^{22}$ cm$^{-2}$, while the lower panel is for the 10$^{23}$ cm$^{-2}$ case. The hatched region marks the common region in all the observed IQRs corresponding to the 5 luminosity bins shown in Figure \ref{fig:corr}.}
    \label{fig:overlap}
\end{figure}

We overlay the IQRs per each AGN continuum luminosity state for each column density case in Figure \ref{fig:overlap}. We mark the overlapping region with hatches in each case that goes on to show that even with an increase in the continuum luminosity and column density, there exists a common family of solutions for the optimal emission of the \hb{} following the flattening trend. This result indicates that although the flattening trend is clear between the \hb{} and AGN continuum luminosities in the two different epochs that has been studied previously \citep{Panda_2022AN....34310091P} and in this work, the trends with respect to increasing column density is different. We will study this effect in detail and explore the possibilities in a forthcoming work.

\section{Conclusions}

We test the connection between the continuum variability in the optical regime (5100\AA) and the corresponding \hb{} response to it for NGC 5548. We perform a multi-component spectral fitting to optical spectra observed in the time range between 48636-49686 (in MJD) and re-affirm the shallowing of the trend between the \hb{} and the corresponding 5100\AA~ AGN continuum luminosities as found in our previous work \citep{Panda_2022AN....34310091P}. We perform a suite of photoionization models assuming constant density, and single cloud approximation using {\sc CLOUDY} in order to infer this observed trend. The flattening of the luminosity-luminosity relation between the continuum and the \hb\ emission line in NGC 5548 is consistent with the presence of a medium of moderate local density, moderate cloud column density whose response diminishes, although there exists a common region in the log \rblr{} - log n parameter space with increasing continuum luminosity. Below, we summarize our findings:

\begin{enumerate}
    \item The saturation of \hb{} luminosity as a function of the rising AGN continuum luminosity in NGC 5548 is evident and follows a 2$^{\rm nd}$ order polynomial.
    \item We obtain consistency between the observed emission and the photoionization models and retrieve the corresponding solutions for the location (\rblr{}) and density (n$_{\rm H}$) for the \hb{}-emitting region as a function of increasing AGN continuum luminosity.
    \item We are able to constrain the local density of the BLR as a function of the rising AGN continuum luminosity for a BLR covering fraction (CF) $\sim$ 20\%.
    \item Even with an increase in the AGN continuum luminosity and column density, there exists a common family of solutions for the \hb{} emission following the flattening trend.
\end{enumerate}

\begin{acknowledgements} SP acknowledges the Conselho Nacional de Desenvolvimento Científico e Tecnológico (CNPq) Fellowships 164753/2020-6 and 300936/2023-0. NB and EB acknowledge the support of the Serbian Ministry of Education, Science and Technological Development, through the contract number 451-03-68/2022-14/200002. \end{acknowledgements} 

\bibliographystyle{mnras}
\bibliography{references}

\end{document}